\documentclass[3p,times]{elsarticle}

\usepackage{ecrc}

\volume{00}

\firstpage{1}

\journalname{Nuclear Physics A}

\runauth{}

\jid{nupha}

\usepackage{amssymb}
\usepackage{graphicx}
\usepackage{comment}

\usepackage[figuresright]{rotating}

\begin{document}

\begin{frontmatter}



\dochead{}

\title{The measurement of non-photonic electrons in STAR}
\author{Olga H\'ajkov\'a for the STAR Collaboration}
\address{Czech Technical University in Prague, Faculty of Nuclear Sciences and Physical Engineering, B\v rehov\'a 7, 11519, Prague 1, Czech Republic}

\begin{abstract}
The measurements of non-photonic electrons (NPE), mainly produced by semileptonic decays of D and B mesons, provide information on heavy quarks production as well as properties of nuclear matter produced in heavy ion collisons. In order to interpret the NPE measurements it is important to determine the relative charm and bottom contribution to the NPE spectrum. In this proceedings we present the measurements of NPE spectra and NPE-hadron azimuthal correlations in p+p collisions at $\sqrt{s}$=200 GeV and at $\sqrt{s}$=500 GeV. NPE-hadron correlations allow extraction of the B decay contribution to the NPE. The B decay contribution is comparable to the contribution from the D meson decay at $\sqrt{s}$=200 GeV at p$_T$ higher than 5 GeV/c, and is about 60\% at $\sqrt{s}$=500 GeV at p$_T$ higher than 5 GeV/c. STAR measured NPE spectrum in p+p collisions as well as relative constribution of bottom decays to the spectrum is consistent with FONLL pQCD calculations. The preliminary results of NPE spectra and NPE-hadron correlations in Au+Au collisions at $\sqrt{s}$=200 GeV is shown.

\end{abstract}


\end{frontmatter}

\section{Introduction}
Due to their large masses, heavy quarks are produced mainly during initial parton-parton interaction at RHIC, and they are good probes to study QCD matter. 
Study of heavy flavor production in p+p collisions is a test of the validity of the perturbative QCD. It is also used as a baseline to study effects of hot and dense nuclear matter from the production of heavy quarks in heavy ion collisions \cite{teory}.   

The nuclear modification factor of non-photonic electrons at p$_T > 6$ GeV/c measured in central Au+Au collisions at $\sqrt{s_{NN}}$=200 GeV is comparable to that of light hadrons \cite{erratum}. Theoretical predictions suggest smaller energy loss of heavy quarks compared with light quarks due to the dead cone effect \cite{dead cone 1} \cite{dead cone 2} \cite{dead cone 3}, if the dominant energy loss process is gluon radiation. In order to address this B and D contribution to the total NPE yield must be quantified. This could be done via non-photonic electrons - charged hadrons azimuthal correlations, taking into account the fact that electrons from D and B decays have different near side correlation shape \cite{shape}.

\vspace{-0.3cm}

\section{Analysis}
Data reported in this proceedings were collected in p+p collisions at $\sqrt{s}$=200 GeV and at $\sqrt{s}$=500 GeV in the years 2005, 2008, and 2009 with High Tower Trigger, and in Au+Au collisions at $\sqrt{s}$=200 GeV in the year 2010 with Minimum Bias. Main detectors used in presented measurements are the Time Projection Chamber (TPC), the main charged particle tracking device in the STAR detector used for particle identification and momentum determination, the Barrel Electromagnetic Calorimeter (BEMC), used for deposited energy measurement, and also as a trigger detector, and the Barrel Shower Maximum Detector (BSMD). 
Electron candidates were identified via specific ionization energy loss from the TPC, the ratio of track momentum to the energy deposited in the BEMC, the BSMD shower profile, and the distance between TPC track projected position at BEMC and reconstructed BEMC cluster position. The obtained inclusive electron sample includes non-photonic electrons, photonic electrons background, and hadron contamination. Non-photonic electrons yield is calculated as:
 
\begin{displaymath} 
N_{NPE} = N_{Inclusive}*\epsilon_{purity} - N_{PHE}/\epsilon_{photonic},
\end{displaymath} 

where $N_{NPE}$ is non-photonic electrons yield, $N_{Inclusive}$ is all identified electrons yield, $\epsilon_{purity}$ is the purity of inclusive electron sample,  $N_{PHE}$ is yield of reconstructed photonic electron background, mainly comes from photon conversion in the detector material and from Dalitz decay of $\pi^0$ and $\eta$ mesons, and $\epsilon_{photonic}$ is photonic electron reconstruction efficiency. This efficiency was determined from embedding simulated gammas and pions into real data. The photonic electron reconstruction efficiency was found to be 0.3 - 0.7 as an increasing function of p$_T$ \cite{clanek_pp}.

\vspace{-0.3cm}

\section{Non-photonic electrons in p+p collisions}

\begin{figure}[h]
  \centering
  \includegraphics[width=7cm]{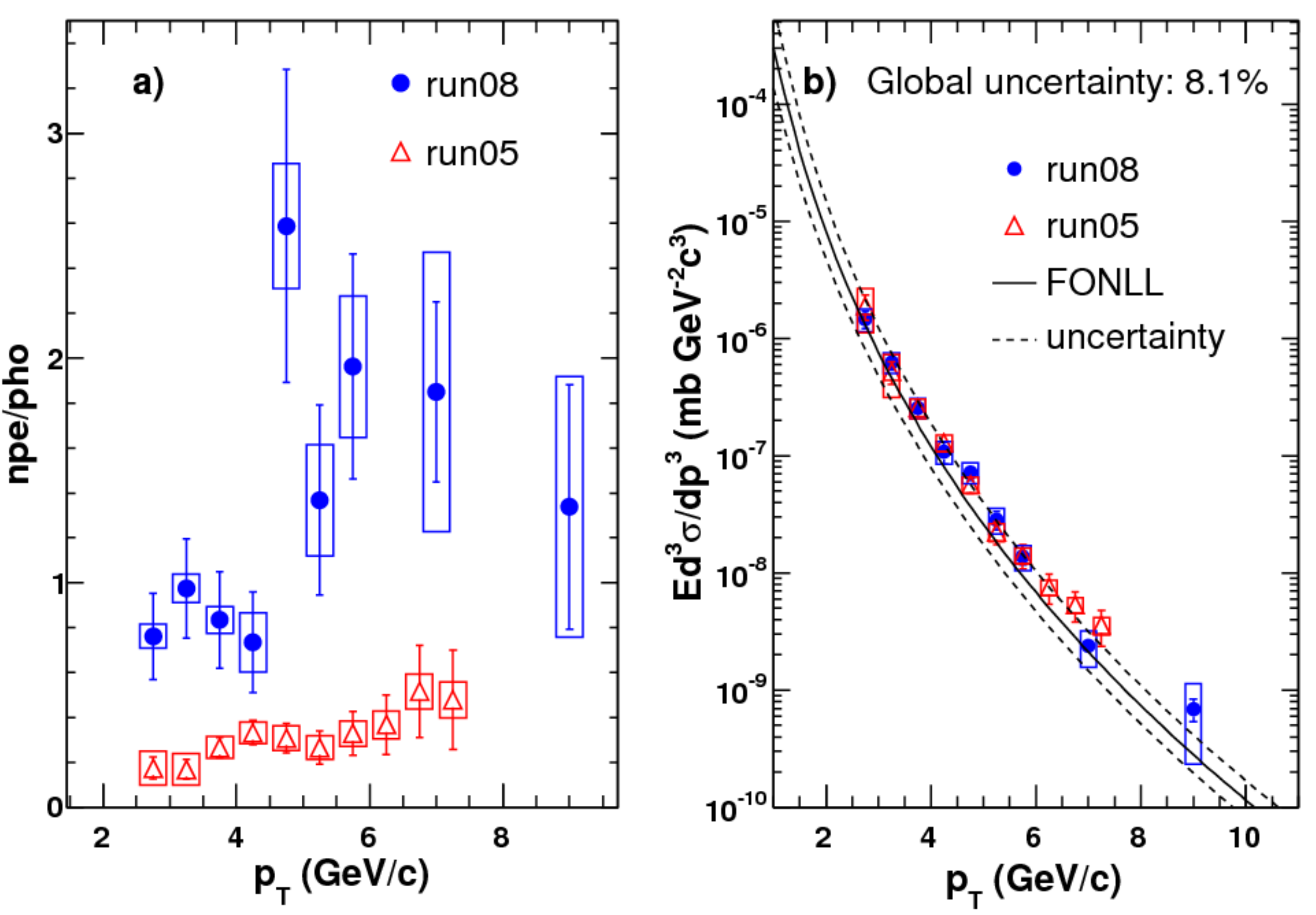}
  \caption{Left: The non-photonic to photonic electrons ratio in p+p collisions at $\sqrt{s}$=200 GeV in the year 2005 (red triangles), and in the year 2008 (blue circles). Right: Non-photonic electron invariant cross section from the years 2005 and 2008 (red triangles and blue circles respectively). The solid line is the FONLL calculation and dashed lines are FONLL uncertainties. \cite{clanek_pp} \cite{FONLL}.}
\label{1-1} 
\end{figure}

\begin{figure}[h]
 \centering
\includegraphics[width=5.3cm]{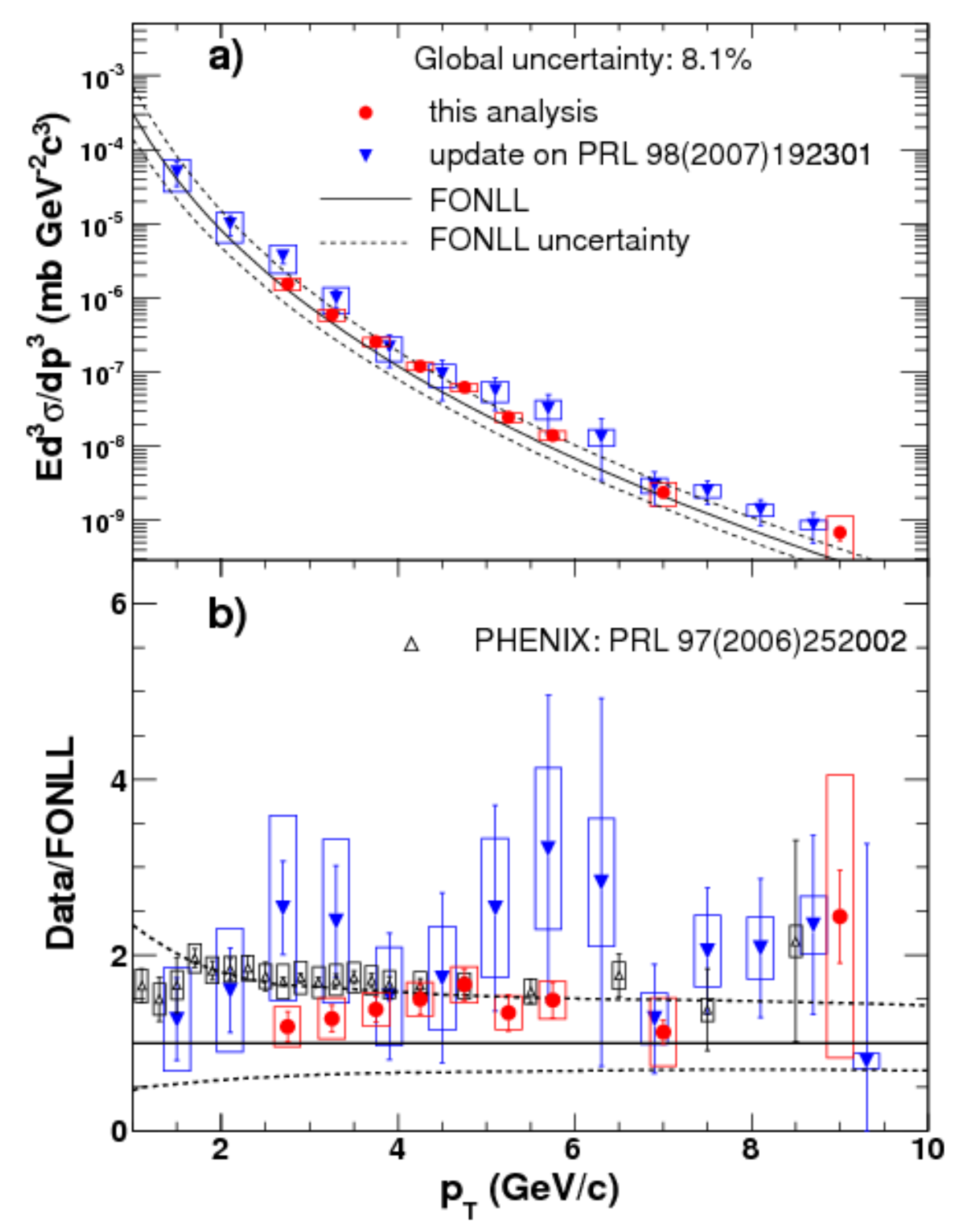}
\caption{Top: Invariant cross section of non-photonic electrons after combining results from the years 2005 and 2008 (red circles). Bottom: The data over FONLL ratio \cite{clanek_pp} \cite{FONLL}.}
\label{1-2}
\end{figure}

Figure \ref{1-1} shows a comparison between results in p+p collisions at $\sqrt{s}$=200 GeV in the years 2005 and 2008. In the year 2005 the STAR detector setup included the  silicon vertex detector in front of the TPC that led to much more gamma conversion background, and consequently to low NPE/PHE ratio, where PHE is photonic electron background. Due to the less material run, the NPE/PHE ratio is much larger in the year 2008 than in the year 2005 - see Figure \ref{1-1} (left). Despite of the large difference in photonic background, results from the years 2005 and 2008 agree with each other - see Figure \ref{1-1} (right). After combining both results, the invariant cross section of non-photonic electron production in p+p collisions was obtained and compared with the FONLL calculation - see Figure \ref{1-2} (up). In Figure \ref{1-2} (down) the data over FONLL ratio are shown. Data from PHENIX and corrected results from the year 2003 are shown there as well. All these results agree with each other \cite{clanek_pp} \cite{erratum}.

\begin{figure}[h]
  \centering
  \includegraphics[width=5.9cm]{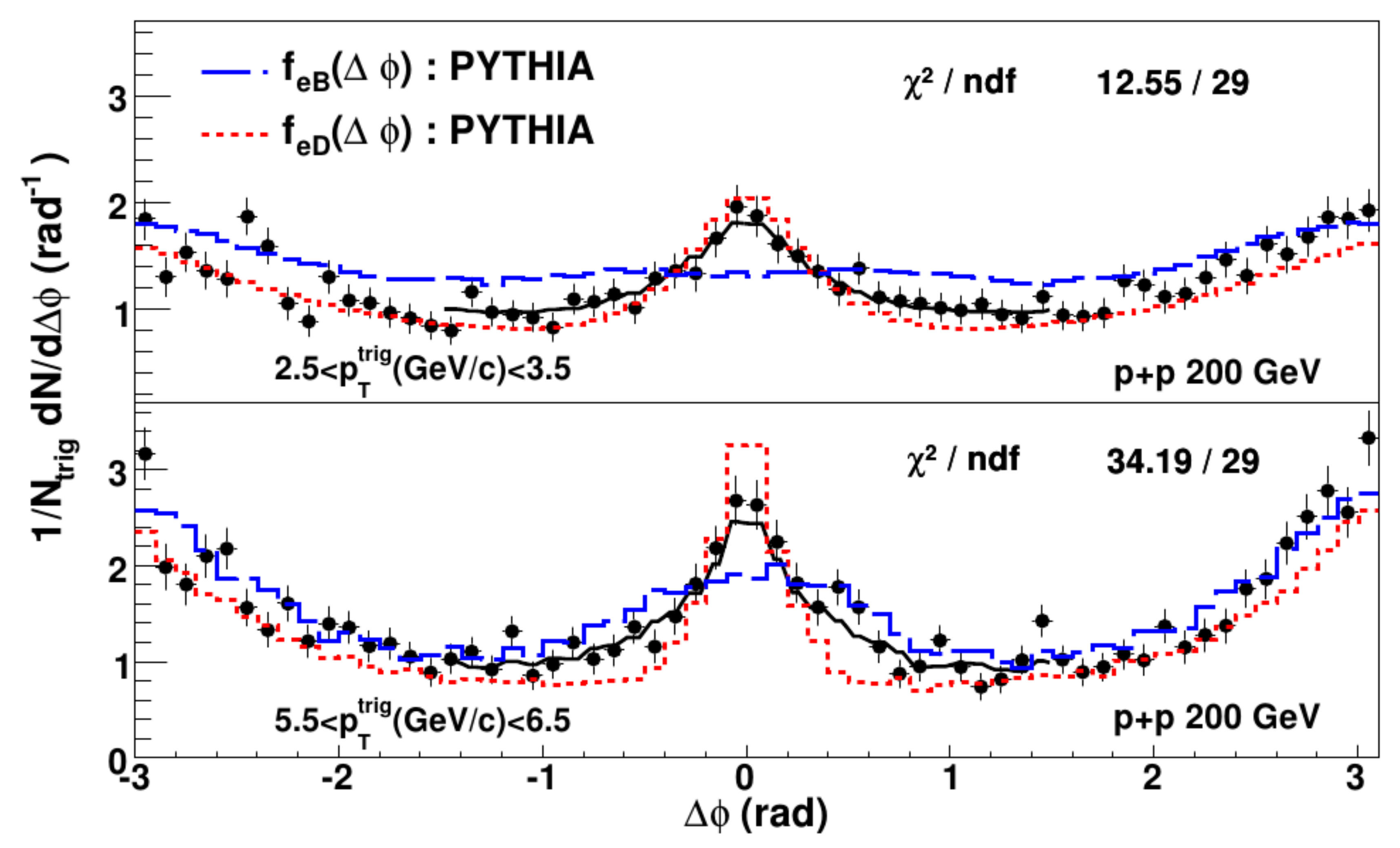}
\includegraphics[width=5.4cm]{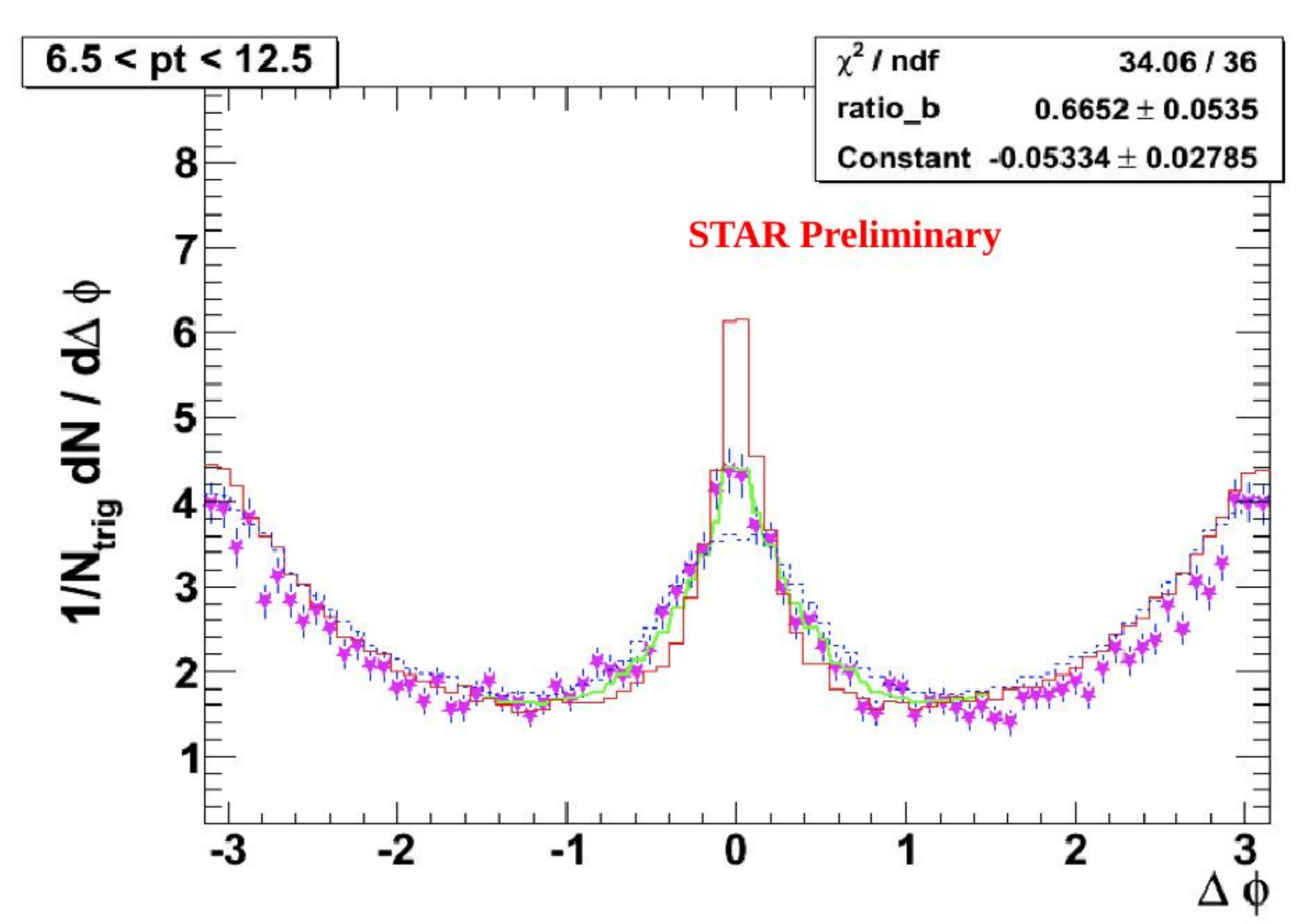}
\includegraphics[width=4.8cm]{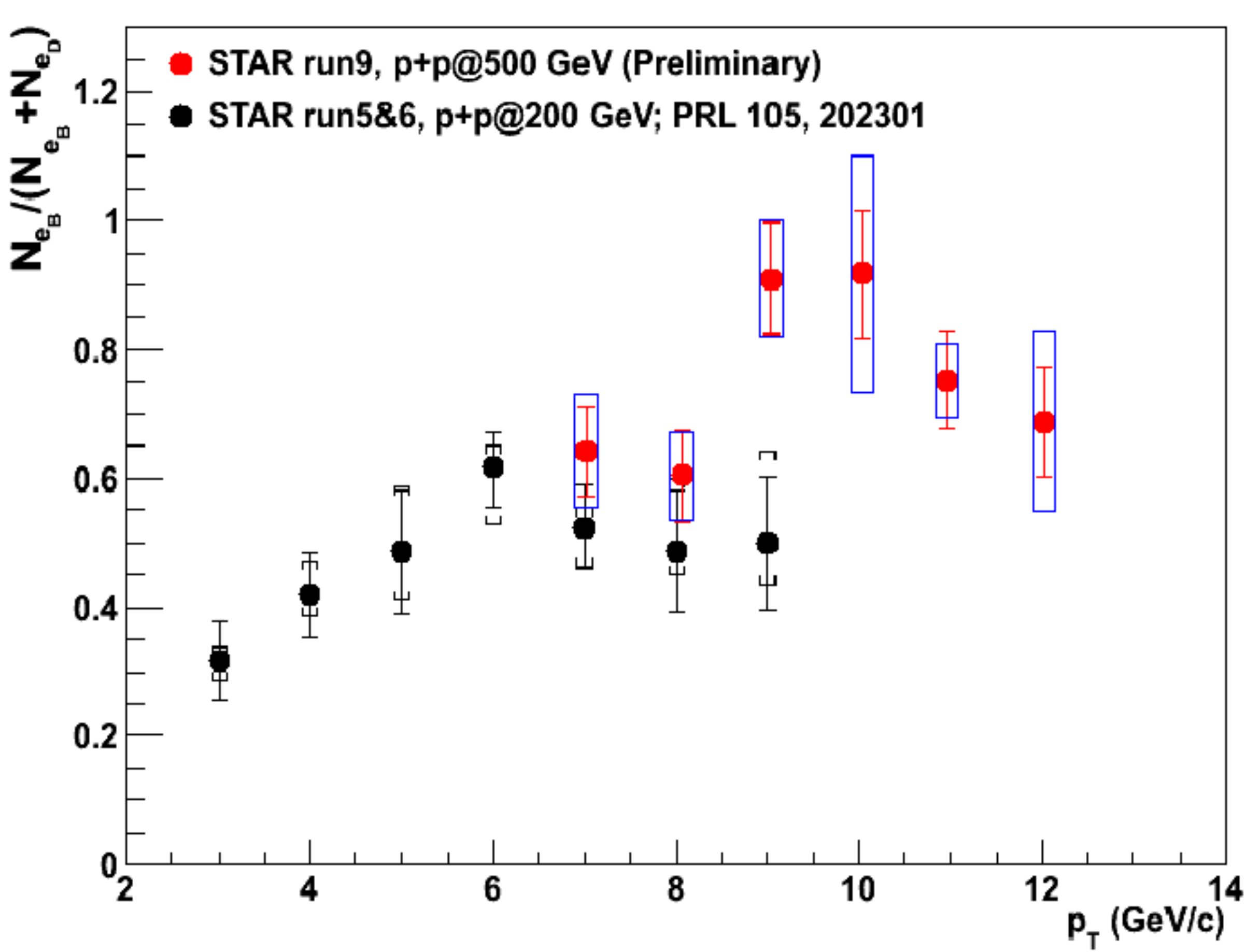}

  \caption{Left/Central: NPE-hadron correlations from data in p+p collisions at $\sqrt{s}$=200 GeV and $\sqrt{s}$=500 GeV (black/magenta dots) compared with PYTHIA simulations of electron(from B/D mesons decays)-hadron correlations (blue dashed and red dotted/solid red lines respectively). The black/green lines are combined fits to the data. Right: The relative bottom contribution to NPE electrons in p+p collisions at $\sqrt{s}$=200 GeV (black dots), and at $\sqrt{s}$=500 GeV (red dots)\cite{correlace}.}
\label{2}
 \end{figure}

NPE originate dominantly from the bottom and charm meson decays. NPE from these two sources could be separated via charged hadron-NPE azimuthal correlations study. The relative B meson contribution to NPE could be obtained by comparing NPE-hadron correlations from data with PYTHIA calculation (Figure \ref{2} left at $\sqrt{s}$=200 GeV, and central at $\sqrt{s}$=500 GeV), fitting the data with PYTHIA shapes for the charm and bottom part. In Figure \ref{2} (right) the relative B contribution as a function of $p_{T}$ at $\sqrt{s}$=200 GeV and $\sqrt{s}$=500 GeV is shown. The B decay contribution increases with p$_T$, and is comparable to the contribution from the D meson decay at p$_T$ higher than 5 GeV/c at $\sqrt{s}$=200 GeV. The ratio of the B contribution to NPE is about 60\% for p+p collisions at 500GeV at high p$_T$. The B contribution is systematically higher at 500 GeV than at 200 GeV in the overlap p$_T$ region.  

Using the information of the relative contribution of the bottom to NPE spectra, it is possible to compare the charm as well as the bottom NPE spectra to FONLL calculations. Spectra and calculations are consistent \cite{clanek_pp}.

\vspace{-0.3cm}

\section{Non-photonic electrons in Au+Au collisions}
\begin{figure}[h]
  \centering
  \includegraphics[width=5cm]{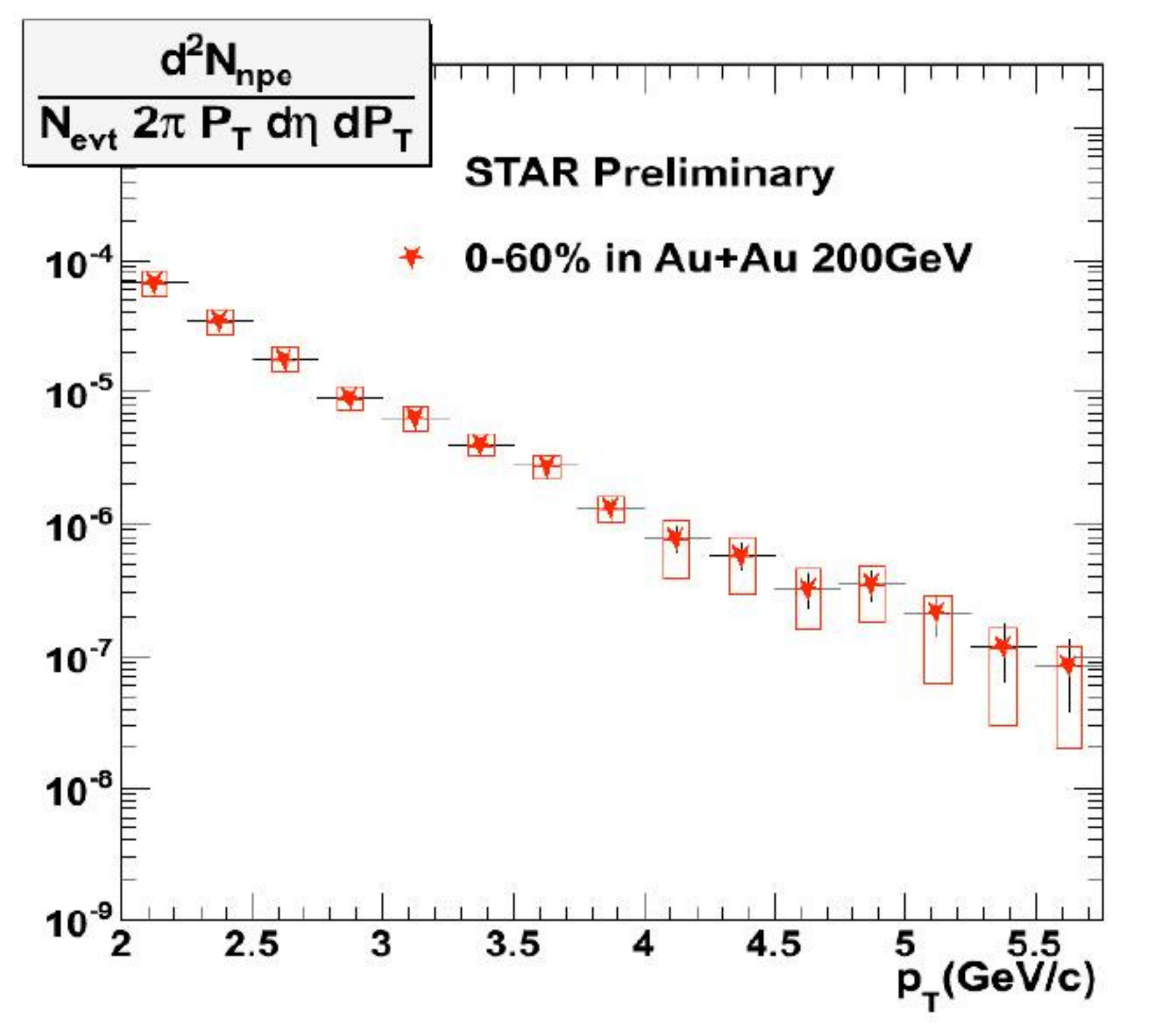}
\includegraphics[width=4.5cm]{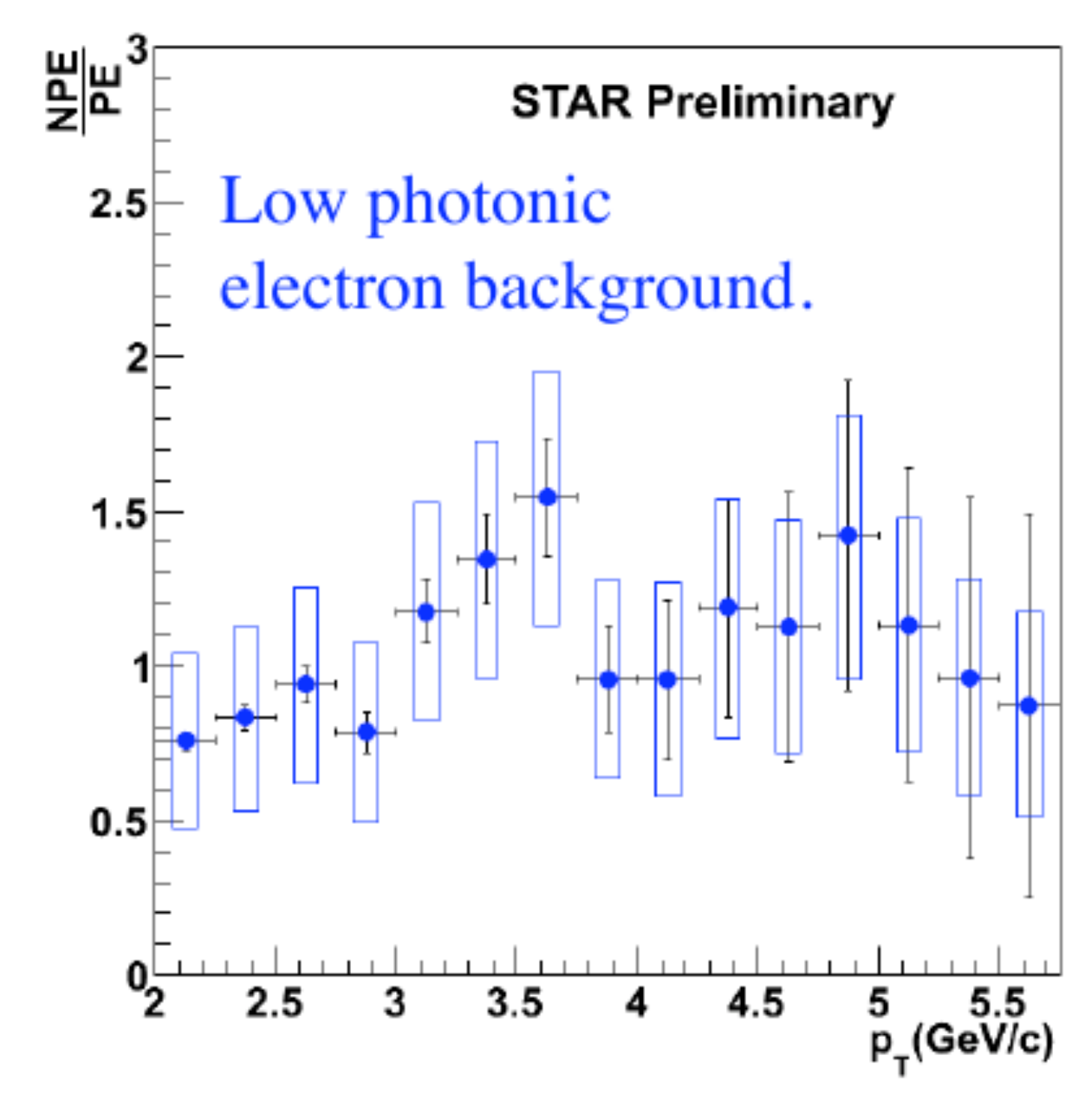}
  \caption{Left: The preliminary result of the non-photonic electron spectrum, for 0-60\% of the most central Au+Au collisions at  $\sqrt{s_{NN}}$=200 GeV. Right: The NPE over PHE ratio as a function of p$_T$.}
\label{3}
 \end{figure}

The preliminary result of non-photonic spectrum in Au+Au collisions at $\sqrt{s_{NN}}$=200 GeV from the year 2010 is plotted in Figure \ref{3} (left). In the right plot there is the non-photonic electrons over photonic electrons ratio as a function of p$_T$. For these results, just a part of data collected in Au+Au collisions in the year 2010 was used. 

\begin{figure}[h]
\centering
\includegraphics[width=3.8cm]{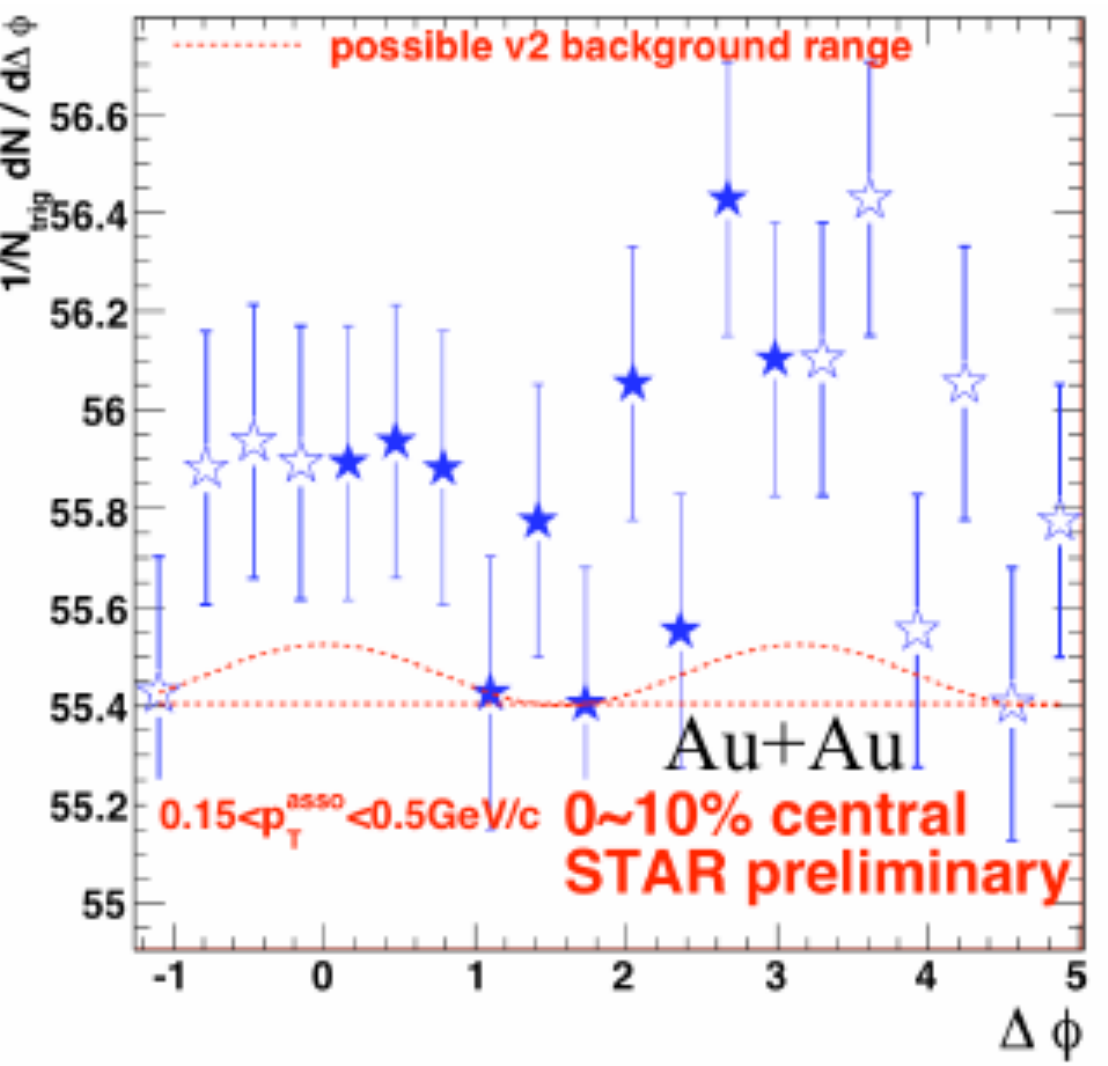}
\includegraphics[width=3.8cm]{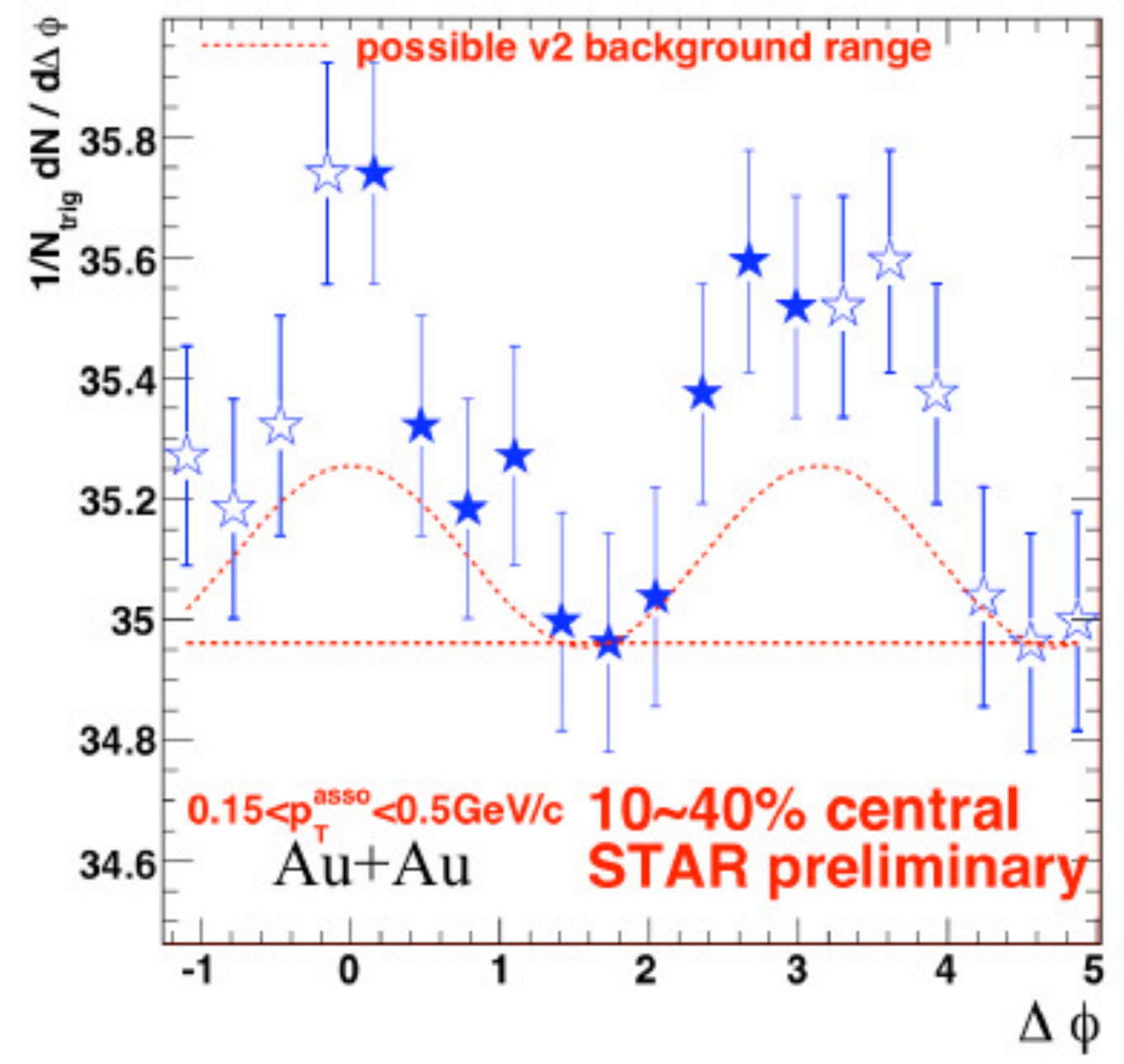}
\includegraphics[width=3.8cm]{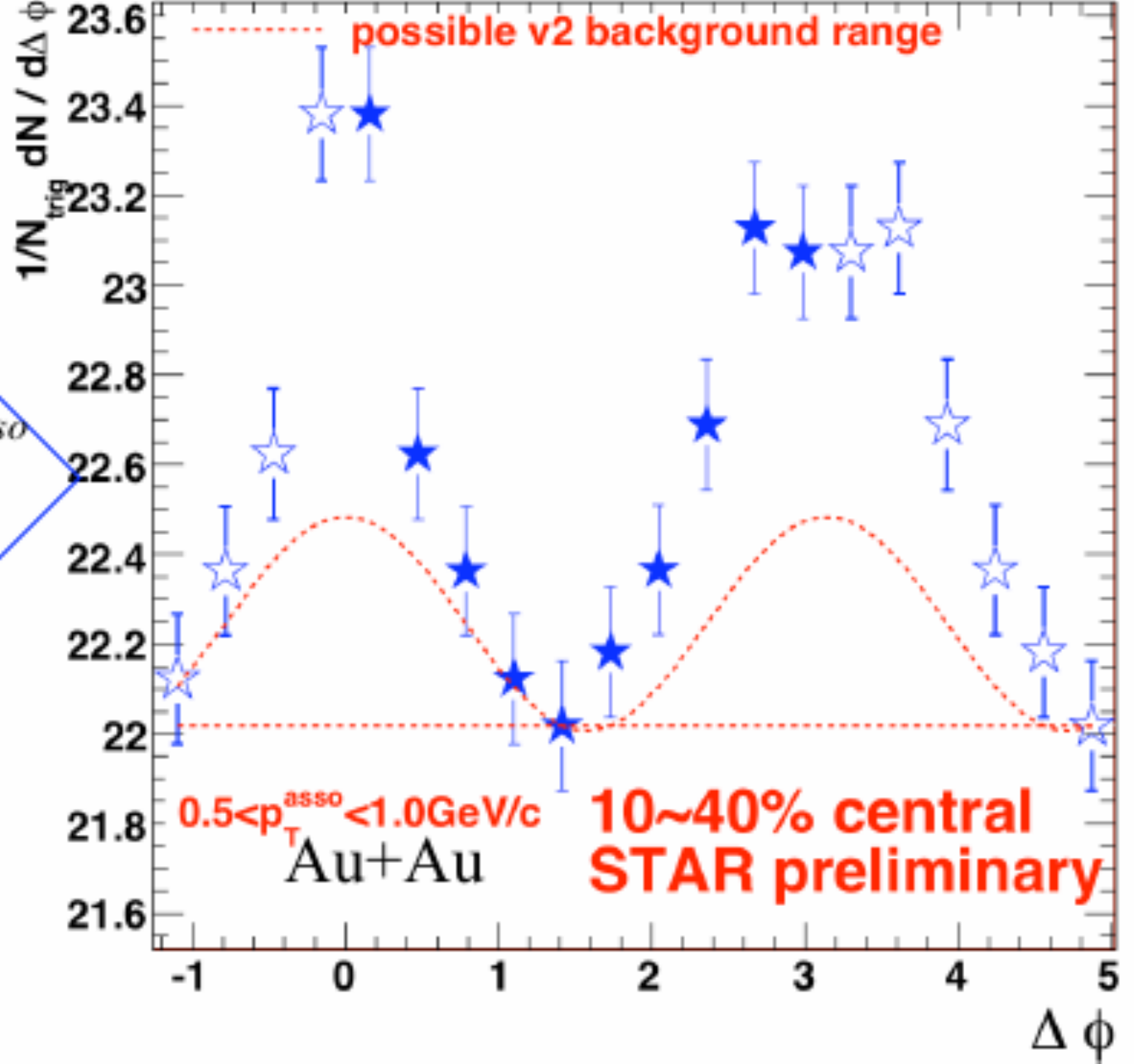}
\caption{Non-photonic-hadron azimuthal corelations in Au+Au collisions at  $\sqrt{s_{NN}}$=200 GeV for 0-10\% central collisions (left), 10-40\% central collisions with associated track 0.15$<p_T^{asso}<$0.5 GeV/c (central), and 10-40\% central collisions with associated track 0.5$<p_T^{asso}<$1 GeV/c (right). Red dotted lines represent upper and lower limits of the estimated NPE elliptic flow $v_2$ background.}
\label{4}
\end{figure}

Non-photonic-hadron correlations in 0-10\% and 10-40\% of the most central Au+Au collisions are plotted in Figure \ref{4}. Correlations in 10-40\% central collisions are plotted with associated tracks with different p$_T^{asso}$. Both the near side and the away side correlations are observed. Due to the fact that the NPE elliptic flow $v_2$ is still under the study, the $v_2$ background is not subtracted yet. In each plot there are two red dotted curves representing the minimum and the maximum of the supposed NPE $v_2$. The minimum possible $v_2$ is estimated to be zero, the maximum is assumed to be the same as hadron $v_2$ \cite{hadronv2}. The NPE elliptic flow $v_2$ was calculated from the electron-event plane correlations. The preliminary result for 10-40\% central Au+Au collisions shows finite $v_2$ \cite{wenqin}.

\section{Summary and Outlook}

In this proceedings, results of non-photonic electrons measurements in p+p collisions at $\sqrt{s}$=200 GeV and $\sqrt{s_{NN}}$=500 GeV and preliminary results of the NPE analysis in Au+Au collisions at $\sqrt{s_{NN}}$=200 GeV from STAR are presented. The preliminary result for p+p collisions at $\sqrt{s}$=500 GeV shows that the B contribution is above 60\% and seems to be larger than in 200 GeV. Two STAR detector upgrades planned for the years 2013 and 2014 will significantly improve the open heavy flavor measurements. Muon Telescope Detector \cite{MTD} will allow muon measurements at mid-rapidity, and Heavy flavor tracker \cite{HFT} will allow measurements of displaced vertices of charm and bottom decays. 

\hspace{0.2cm}

This work was supported by grant INGO LA09013 of the Ministry of Education, Youth and Sports of the Czech Republic, and by the Grant Agency of the Czech Technical University in Prague, grant No. SGS10/292/OHK4/3T/14.   

\hspace{0.5cm}


\begin{thebibliography}{00}
\bibitem{teory} A. D. Frawley, T. Ullrich and R. Vogt, Phys. Rept. 462, 125 (2008).
\bibitem{erratum} B. I. Abelev et al. [STAR Collaboration], Phys.Rev.Lett. 98, 192301 (2007); Erratum-ibid. 106, 159902 (2011).
\bibitem{dead cone 1}M. Djordjevic, M. Gyulassy, S. Wicks, Phys. Rev. Lett. 94, 112301 (2005).
\bibitem{dead cone 2}N. Armesto et al., Phys. Rev. D71, 054027 (2005).
\bibitem{dead cone 3}Y. L. Dokshitzer, D. E. Kharzeev, Phys. Lett. B519, 199 (2001).
\bibitem{shape}J. Adams et al. [STAR Collaboration], Phys. Rev. Lett. 105, 202301 (2010).
\bibitem{clanek_pp}H. Agakishiev et al. [STAR Collaboration], Phys. Rev. D83, 052006,(2011).
\bibitem{FONLL}M. Cacciari, P. Nason and R. Vogt, Phys. Rev. Lett. 95, 122001 (2005).
\bibitem{correlace}M. M. Aggarwal et al. [STAR Collaboration], Phys. Rev. Lett. 105, 202301 (2010).
\bibitem{hadronv2}J. Adams et al. [STAR Collaboration], Phys. Rev. C 72, 014904 (2005).
\bibitem{wenqin}Wenqin Xu [STAR Collaboration], arXiv:1106.6020v1 (2011).
\bibitem{MTD} L. Ruan, G. Lin (Yale U.), Z. Xu, K. Asselta, H.F. Chen, W. Christie, H. J. Crawford, J. Engelage, G. Eppley, C. Li et al. [STAR Collaboration], J. Phys. G G36, 095001, (2009).
\bibitem{HFT}Spiros Margetis [STAR Collaboration], Nucl. Phys. Proc. Suppl. 210-211 (2011). 
\end{thebibliography}
\end{document}